# A Gapped Phase in Semimetallic $T_d$-WTe$_2$ Induced by Lithium Intercalation

Mengjing Wang[1,2], Aakash Kumar[1,2], Hao Dong[2,3], John M. Woods[1,2], Joshua V. Pondick[1,2], Shiyu Xu[1,2], Peijun Guo[2,3], Diana Y. Qiu[1,2], Judy J. Cha[1,2*]

[1] Department of Mechanical Engineering and Materials Science, Yale University, New Haven, CT 06511, USA.

[2] Energy Sciences Institute, Yale West Campus, West Haven, CT 06516, USA.

[3] Department of Chemical and Environmental Engineering, Yale University, New Haven, CT 06511, USA.

* judy.cha@yale.edu

**Abstract**

The Weyl semimetal WTe$_2$ has shown several correlated electronic behaviors, such as the quantum spin Hall effect, superconductivity, ferroelectricity, and a possible exciton insulator state, all of which can be tuned by various physical and chemical approaches. Here, we discover a new electronic phase in WTe$_2$ induced by lithium intercalation. The new phase exhibits an increasing resistivity with decreasing temperature and its carrier density is almost two orders of magnitude lower than the carrier density of the semi-metallic $T_d$ phase, probed by *in situ* Hall measurements as a function of lithium intercalation. Our theoretical calculations predict the new lithiated phase to be a charge density wave (CDW) phase with a bandgap of ~ 0.14 eV, in good agreement with the *in situ* transport data. The new phase is structurally distinct from the initial $T_d$ phase, characterized by polarization angle-dependent Raman spectroscopy, and large lattice distortions close to 6 % are predicted in the new phase. Thus, we report the first experimental evidence of CDW in $T_d$-WTe$_2$, projecting WTe$_2$ as a new playground for studying the interplay between CDW and superconductivity. Our finding of a new gapped phase in a two-dimensional (2D) semi-metal also demonstrates electrochemical intercalation as a powerful tuning knob for modulating electron density and phase stability in 2D materials.

**Main**

Two-dimensional (2D) orthorhombic tungsten ditelluride ($T_d$-WTe$_2$) is a type-II Weyl semimetal in bulk and a topological insulator as a monolayer, stimulating studies of quantum electronic phases and search for topological superconductivity[1-4]. Superconductivity with $T_c$ of 580 mK has been realized in monolayer $T_d$-WTe$_2$ by electrostatic doping at an electron density of $5\times10^{12}$ e/cm$^2$ [1, 2], and $T_c$ was raised to ~ 2.6 K by chemical intercalation of potassium in bulk $T_d$-WTe$_2$ where the potassium is expected to donate charge to WTe$_2$ [5]. With electron doping, a charge density wave (CDW) may emerge in WTe$_2$ as several 2D chalcogenides, such as TaS$_2$, possess superconductivity as well as CDWs that are modulated by electron density[6-9]. Recent theoretical papers indeed predict that a high level of electron doping of over $10^{14}$ e/cm$^2$ in monolayer $T_d$-WTe$_2$ can introduce lattice distortions and form a CDW phase[10-12], neighboring the superconducting dome at $10^{13}$ e/cm$^2$. However, no experimental reports of a CDW phase in WTe$_2$ are found in the literature.

Here, we use lithium intercalation to modulate the carrier density of WTe$_2$ beyond the charge doping range achievable by ionic liquid gating and discover a new electronic phase in WTe$_2$, which we identify as a CDW phase. With controllable lithium intercalation and deintercalation, we observe a reversible structural and electronic phase transition as evidenced by *in situ* optical, Raman, and electrical transport characterization throughout the intercalation. Structurally, the new phase shows distinct Raman peaks with crystal symmetry different from the T$_d$ symmetry, in agreement with the T$_d$' phase recently reported by Muscher *et al* [13]. Electronically, despite the significant electron doping by lithium, the new phase exhibits a reduction of electron carrier density by almost two orders of magnitude compared to the pristine semi-metallic WTe$_2$ and increasing resistance with decreasing temperature, suggesting a bandgap opening. Our *ab initio* band structure computed within density functional theory (DFT) shows that the lithiated T$_d$' phase develops an indirect bandgap of 0.14 eV and a direct bandgap of 0.66 eV, consistent with the experimentally observed semiconductor-like electrical behavior. Furthermore, calculations of the phonon dispersion reveal that the T$_d$ phase exhibits a phonon softening with lithium intercalation at the level of the experimental phase transition and at wave vectors roughly consistent with the formation of a 2×2 folding of the Brillouin zone. Concomitantly, lithiation stabilizes the T$_d$' phase, which is a 2×2 superlattice of the T$_d$ phase involving a distortion of the W chain along *a*-axis where the W-Te-W bond lengths are shortened and elongated in their successive arrangements in contrast to the identical W-Te-W bond lengths in the T$_d$ phase. Overall, the *a* and *c* lattice parameters are strained by 4.9 % and 5.84 % respectively when WTe$_2$ changes from the T$_d$ to the T$_d$' phase. Thus, we report the first experimental evidence of a gapped CDW phase in WTe$_2$ enabled by lithium intercalation.

**Lithium intercalation induced phase transition**

For lithium intercalation, we construct a WTe$_2$ flake electrochemical cell using liquid or polymer gel electrolyte with capabilities of *in situ* Raman and *in situ* transport measurements (**Figure 1a**, Methods). The exfoliated WTe$_2$ flakes have a lateral size distribution of 20 ~ 50 μm and a thickness range of 30 ~ 100 nm (> 50 layers) (**Figure S1**, Supplementary Information). By controlling the electrochemical intercalation voltage ($V_{EC}$) between the WTe$_2$ flake (cathode) and lithium metal (anode), lithium ions can be controllably inserted into the van der Waals gaps of the WTe$_2$ flake (**Figure 1b**). Lower values of $V_{EC}$ represent more lithium ions intercalated into WTe$_2$ flakes.

A new structural phase emerges with distinct Raman modes at $V_{EC}$ of 0.8 V *vs* Li$^+$/Li in a polymer cell (**Figure 1c, d, Figure S2,** Supplementary Information) or at $V_{EC}$ of 1.2 V *vs* Li$^+$/Li in a liquid cell (**Figure S3**, Supplementary Information). In the low frequency region (< 50 cm$^{-1}$) of the Raman spectrum where inter-layer vibration modes are active, the shear mode of the T$_d$ phase at ~ 8 cm$^{-1}$ [14] splits into two peaks at 6 cm$^{-1}$ and 13 cm$^{-1}$, as illustrated in **Figure 1c**. In the high frequency region (> 50 cm$^{-1}$) where intralayer phonon modes are populated, several Raman modes with distinct peak positions appear (**Figure 1d**). These new Raman peaks do not match those observed in the initial T$_d$, 1T' or 2H of WTe$_2$ (**Table S1**, Supplementary Information). The Raman spectrum of the new phase has two key characteristics: increased number of Raman active modes compared to the T$_d$ phase and no discernible patterns in the peak shifts of the Raman modes relative to the T$_d$ phase. Therefore, the new Raman modes are not a result of strain or electron doping, but a manifestation of phase transition to a new phase with atomic configurations distinct from the T$_d$ phase. Recently, Muscher *et al.* reported a new lithiated T$_d$'-

WTe$_2$ phase whose Raman modes match our experimental data[13] (**Table S1**, Supplementary Information). Therefore, we identify the new lithiated phase in this work to be the same as that reported by Muscher *et al.* and henceforth refer to it as the lithiated T$_d$' phase (T$_d$' when referring to the structure without any Li atoms). Optically, when WTe$_2$ is in the lithiated T$_d$' phase, dark streaks appear along the crystallographic axis *b*, perpendicular to the direction of the W-W chain, as depicted in the inset of **Figure 1c**. A closer scrutinization of the streaks *via* scanning electron microscopy (**Figure S4**, Supplementary Information) indicates them to be wrinkles, suggesting the cause of the dark streaks to be an anisotropic strain induced during the phase transition from T$_d$ to T$_d$'.

A representative evolution of the high frequency Raman peaks as $V_{EC}$ is lowered from open circuit voltage (OCV) to 0.4 V *vs* Li$^+$/Li in a polymer cell is shown in **Figure 2b** with corresponding optical images in **Figure 2a**. Three distinct stages take place in the course of lithium intercalation. First, WTe$_2$ remains in T$_d$ phase when OCV > $V_{EC}$ > 0.8V *vs* Li$^+$/Li with negligible changes in the Raman spectra and the optical contrast. In the second stage, when 0.8 V > $V_{EC}$ > 0.4 V *vs* Li$^+$/Li, WTe$_2$ is in the new lithiated T$_d$' phase characterized by the distinct Raman modes and dark streaks appearing along the *b* axis (**Figure 2a**). Lastly, when $V_{EC}$ < 0.4 V *vs* Li$^+$/Li, WTe$_2$ becomes amorphous, supported by the featureless Raman spectrum and a yellowish hue color change in WTe$_2$.

The resistance of the WTe$_2$ flake undergoing the phase transition was simultaneously tracked by two terminal measurements of the same device (**Figure 2c**). First, the resistance does not change much while WTe$_2$ remains in the T$_d$ phase (grey shaded region). While $V_{EC}$ is at 0.8 V *vs* Li$^+$/Li, the resistance starts to increase from 386 Ω at ~ 2000 s to 1015 Ω at ~ 3000 s. This resistance increase is surprising as lithium intercalation in transition metal dichalcogenides (such as $MoS_2$) generally leads to a decreased resistivity[15] due to the extra electrons donated by Li. When $WTe_2$ is in the lithiated $T_d'$ phase with the distinct Raman modes ($V_{EC}$ = 0.8 V *vs* $Li^+$/Li between 3000 s to 10000 s, pink shaded region), the resistance continues to increase and eventually saturates at 7032 Ω, a nearly twenty times increase in resistance from the pristine $T_d$ phase. The saturated resistance indicates a completion of the phase transition. At $V_{EC}$ of 0.4 *vs* $Li^+$/Li, $WTe_2$ starts to break down and becomes amorphous, accompanied with an irreversible sharp increase in resistance (purple shaded region).

For further characterization of the crystal structure and the electronic band structure of the lithiated $T_d'$ phase, we performed *in situ* angle-resolved Raman spectroscopy and *in situ* magneto-transport measurements, which we discuss in the following sections.

***In situ* angle-resolved Raman spectroscopy of the lithiated $T_d'$ phase**

To characterize the crystal symmetry of the new phase, we performed *in situ* angle-resolved Raman spectroscopy, as presented in **Figure 3** (full spectra in **Figure S5**, Supplementary Information). As a function of the angle $\theta$ between the polarization direction of the incident laser and the vertical dashed line shown in **Figure 3a**, we acquired the angle dependent Raman spectra of the pristine $T_d$ phase (**Figure 3b**) and the lithiated $T_d'$ phase (**Figure 3e**). The lithiated $T_d'$ phase has more Raman active modes, albeit with weaker intensities, than $T_d$ in the range from 50 to 220 $cm^{-1}$, suggesting a lower crystal symmetry of the $T_d'$ phase. Additionally, the symmetry of the prominent Raman modes is two-fold in the lithiated $T_d'$ phase, such as the peaks centered at 80 and 210 cm$^{-1}$, while the symmetry of the Raman modes is mostly four-fold in the T$_d$ phase, e.g., peaks at 75, 130, and 214 cm$^{-1}$. The pole graphs of the Raman modes near 210 cm$^{-1}$ clearly show the four-lobed and two-lobed nature of the Raman modes in T$_d$ and lithiated T$_d$' phase, respectively (**Figure 3c** and **Figure 3f**).

*In situ* **magneto-transport measurements during lithium intercalation**

The twenty-fold resistance increase of WTe$_2$ from the initial T$_d$ phase to the lithiated T$_d$' phase suggests there may be a significant change in the electronic band structure of WTe$_2$ in the new T$_d$' phase. To answer this, we carried out *in situ* magneto-transport measurements of a WTe$_2$ flake as a function of Li intercalation.

Prior to intercalation, magneto-transport of a pristine WTe$_2$ flake was measured at 2 K, which showed clear Shubnikov-de Hass oscillations in high magnetic fields (**Figure S6**, Supplementary Information), suggesting a high crystal quality of WTe$_2$. Fitting the $R_{xx}$ and $R_{xy}$ to a two-band transport model[16, 17] showed that the pristine WTe$_2$ flake has an almost equally compensated electron and hole density at 2 K ($n_e$ = 2.11×10$^{20}$ cm$^{-3}$; $n_h$ = 1.97×10$^{20}$ cm$^{-3}$; **Figure S6**, Supplementary Information). The nearly balanced electron and hole population indicates that the Fermi level ($E_F$) is close to zero in pristine WTe$_2$ intersecting the conduction band and valence band equally at 2 K, as predicted by DFT[16] and visualized in **Figure 4a**. At 300 K, transport data of the pristine WTe$_2$ flake suggests that electrons are populated more than holes (**Figure S7 f-g**, Supplementary Information) and thereby conduction is primarily carried out by electrons at room temperature.

During intercalation, both $R_{xx}$ and $R_{xy}$ were measured in the magnetic field of 3 T at 300 K as $V_{EC}$ was swept from OCV to 0.8 V, as displayed in **Figure 4c-d**. In accordance with the Raman observation that new Raman modes from $T_d'$ emerge when $V_{EC}$ stays at 0.8 V for 30 mins, a red vertical line in **Figure 4c-d** marks the phase transition point from $T_d$ to lithiated $T_d'$ phase. When WTe$_2$ stays in $T_d$ during OCV > $V_{EC}$ > 0.8 V *vs* Li$^+$/Li, $R_{xx}$ remains almost unchanged and decreases marginally at 1.0 V *vs* Li$^+$/Li from 32 Ω to 28 Ω, likely due to electron doping from Li. At $V_{EC}$ of 0.8 V *vs* Li$^+$/Li, $R_{xx}$ rises immediately and saturates at 2620 Ω after 50 mins at 0.8 V *vs* Li$^+$/Li, indicating a completion of phase transition from $T_d$ to lithiated $T_d'$ phase. The increase of $R_{xx}$ measured with four probes agrees with the two-terminal measurement shown in **Figure 2c**.

Unlike the steady increase of $R_{xx}$ at $V_{EC}$ = 0.8 V *vs* Li$^+$/Li, the Hall resistance $R_{xy}$ shows a plateau that lasts about 30 mins before it starts to increase. According to the *in situ* Raman data, WTe$_2$ stays as $T_d$ phase during the first 30 mins at $V_{EC}$ = 0.8 V. Therefore, the plateau region of $R_{xy}$ suggests a potential intermediate WTe$_2$ phase, which does not have drastic lattice rearrangements that can be detected by Raman spectroscopy but possesses a distinct electron band structure from that of $T_d$. After the plateau, $R_{xy}$ increases and eventually saturates at 45 Ω in the new $T_d'$ phase. After removing the contribution from $R_{xx}$ due to the imperfect alignment of the Hall electrodes (see Methods), the plateau region persists in the corrected $R_{xy}$, but the sign changes from positive to negative in the new $T_d'$ phase, suggesting a hole-dominated transport (**Figure S8**, Supplementary Information). **Figure 4e** shows the Hall carrier density ($n_{Hall}$) determined from the corrected $R_{xy}$. When OCV > $V_{EC}$ > 0.8 V *vs* Li$^+$/Li, $n_{Hall}$ remains unchanged at 7.6×10$^{15}$ cm$^{-2}$ as WTe$_2$ is in the $T_d$ phase. The spikes seen in $n_{Hall}$ correspond to the regions where the differences of $R_{xy}$ values between 3 T and 0 T are very small, and therefore are artifacts and not physically meaningful. At the onset of $V_{EC}$ = 0.8 V, $n_{Hall}$ falls considerably to $9.1 \times 10^{14}$ cm$^{-2}$ and remains at this density. After ~ 30 mins at $V_{EC}$ = 0.8 V, WTe$_2$ is in the lithiated T$_d$' phase and $n_{Hall}$ changes the sign with a further reduced hole carrier density of $1.9 \times 10^{14}$ cm$^{-2}$.

The observed characteristics of $R_{xx}$ and $R_{xy}$ are reproduced in the second round of lithium intercalation (**Figure S7 a-c**, Supplementary Information) and in a different Hall bar device (**Figure S9**, Supplementary Information), confirming the reversibility of the phase transition and ruling out the possibilities of electrical contact degradation for the observation. We also confirm that the electrical contacts remain unaffected by intercalation as the magneto-transport behaviors remain unchanged between pristine state and deintercalated state (**Figure S7 f-g**, Supplementary Information). Therefore, the increase of resistance and accompanying reduction of carrier density exclusively originate from the change in the electronic band structure as WTe$_2$ changes to the new lithiated T$_d$' phase from T$_d$.

The reduced carrier density and the change of the carrier type suggests a potential gap opening in initially semimetallic WTe$_2$. To test this hypothesis, temperature-dependent $R_{xx}$ were acquired at pristine state (**Figure S10,** Supplementary Information) as well as at various stages of intercalation by first halting the intercalation through rapid quenching to 200 K, as shown in **Figure 4b**. At OCV > $V_{EC}$ > 0.8 V *vs* Li$^+$/Li, lithiated WTe$_2$ remains in the T$_d$ phase and $R_{xx}$ *vs* T curves show a metallic transport with decreasing resistance at lower temperatures. In contrast, at $V_{EC}$ ≤ 0.8 V *vs* Li$^+$/Li, the lithiated WTe$_2$ behaves like a semiconductor with resistance increasing with decreasing temperature. This suggests that the new lithiated T$_d$' phase has a bandgap, in contrast to the starting semimetallic T$_d$ phase, as illustrated in **Figure 4a**. We note that the temperature range was limited to 150 K for transport measurements due to the sudden volume change of the polymer electrolyte at that temperature, which breaks the devices.

We carried out DFT calculations of the electronic band structure of the $T_d$ and $T_d'$ phases (**Figure 5a-b**) to elucidate the origin of the dramatic differences in the electron transport of these phases. The lithiated $T_d'$ phase has a simple monoclinic[13] conventional unit cell with twice the number of atoms of the orthorhombic $T_d$ unit cell, as illustrated in **Figure 5a-b**. Detailed crystallographic information of $T_d$ and lithiated $T_d'$ is summarized in **Table S2** and **Table S3**, Supplementary Information. The computed band structure for pristine $T_d$ phase in **Figure 5c** shows that the electron pocket is slightly larger than the hole pocket, consistent with our magneto-transport data of pristine $T_d$ state. Moreover, the band structure of the lithiated $T_d'$ phase in **Figure 5d** confirms that there is a gap opening in the lithiated $T_d'$ phase with a direct gap of 0.66 eV and an indirect gap of 0.14 eV along the Γ-X direction. The proximity of the Fermi level to the valence band maximum further suggests a hole-dominant transport behavior in lithiated $T_d'$ phase, which confirms the finding of electron to hole transport transition in our *in situ* magneto-transport measurements.

**Mechanism of the phase transition and formation of CDWs**

The crystal structure of the lithiated $T_d'$ phase was recently discovered using single crystal X-ray crystallography by Muscher *et al* [13]. Remarkably, the same crystal structure was theoretically predicted to be a 2×2 CDW in monolayer $WTe_2$ under electron doping at a level of 0.5 electrons per unit cell (9.04 ×$10^{13}$ e/$cm^2$) [10, 11]. This strongly suggests that the new phase we observe coincides with the predicted 2×2 CDW and that the CDW formation is driven by the extremely high levels of electron doping from Li intercalants.

To confirm the CDW nature of the new phase in bulk and explore the mechanism driving the phase transition, we computed the phonon dispersion of the lithiated $T_d$ phase in bulk as shown in **Figure 6**. Our Bader charge[18] analysis suggests that each Li atom dopes ~ 0.85 e to the Te atoms. Accordingly, we computed the phonon dispersion of the lithiated $T_d$ phase for two concentrations of electronic doping: 1 Li atom/unit cell ($T_d$-Li$_{0.25}$WTe$_2$) in **Figure 6b** and implicit doping at a concentration equivalent to the level of doping by Li in the lithiated $T_d'$ phase ($T_d$-Li$_{0.5}$WTe$_2$) in **Figure S11**, Supplementary Information. The phonon dispersion of the pristine $T_d$ phase in **Figure 6a** shows a stable structure. Upon introduction of 1 Li atom per unit cell, the phonon branch along the X-S-Y path in the Brillouin zone becomes soft, suggesting the instability of the lattice and the possible formation of a CDW, similar to the picture of the CDW phase diagram predicted in monolayer WTe$_2$[10]. A similar softening is also observed in the case of the implicitly doped $T_d$ in **Figure S11**, Supplementary Information, indicating that high level of electron doping alone leads to lattice instability. On the contrary, the lithiated $T_d'$ phase ($T_d'$-Li$_{0.5}$WTe$_2$) shows a stable phonon dispersion in **Figure 6c** while the removal of Li atoms in the $T_d'$ phase softens the phonon modes along the X-H$_1$-C-H-Y path in **Figure 6d**. Energetically, the phase transition from $T_d$ to the $T_d'$ is driven by the greater negative formation energy of $T_d'$ (2×2 CDW) than the lithiated $T_d$-WTe$_2$ at the same Li concentration. Based on our DFT calculations, the change in enthalpy in going from $T_d$-WTe$_2$ to lithiated $T_d'$-Li$_{0.5}$WTe$_2$ is -0.62 eV, whereas it is -0.36 eV going from $T_d$ to lithiated $T_d$-Li$_{0.5}$WTe$_2$.

**Conclusion**

A new gapped phase is discovered in semi-metallic $T_d$-WTe$_2$, stabilized by an exceptionally high level of electron doping from lithium intercalation. We identified the structure of the new phase to be a lithiated $T_d'$ phase by *in situ* angle-resolved Raman spectroscopy. A bandgap opening in the new phase is suggested by *in situ* transport measurements that show an increasing resistance with decreasing temperature and a much-reduced carrier density. Our DFT calculations show that this new lithiated $T_d'$ phase has an indirect gap of 0.14 eV and a direct bandgap of 0.66 eV. Furthermore, the structure of the lithiated $T_d'$ phase is identical to the unit cell of the predicted 2×2 CDW in WTe$_2$ monolayer by electron doping[10], and phonon dispersion calculations show that lithiation induces phonon softening in the $T_d$ phase at various electron doping levels. Therefore, we identify this new lithiated $T_d'$ phase to be a 2×2 CDW in WTe$_2$, landing at the high electron doping end in the electronic phase diagram of WTe$_2$, past the superconductivity phase of WTe$_2$ [11].

Our discovery of the CDW phase in type II Weyl semimetal WTe$_2$ not only enriches the phase diagram of WTe$_2$, but also renders WTe$_2$ a versatile material platform to study the interplay between superconductivity and CDW. We envision that the exceptional electron doping using lithium intercalation can also introduce CDW in other 2D transition metal dichalcogenide systems that are not known to host CDW in their intrinsic state, such as 1T'-MoTe$_2$. Thus, the greatly expanded electronic and structural phase diagram of 2D materials accessible using lithium intercalation will stimulate the research for novel quantum phases, such as topological superconductivity. Additionally, the sharp switch of electrical resistance of layered WTe$_2$ by controllable tuning of electrochemical intercalation voltage projects prospective applications in resistive memory.

**Acknowledgements**

M. W. and J. J. C. gratefully acknowledge the support from the Moore Foundation under the EPiQS Synthesis Investigator Award program. The electrochemical cell was developed with the support from the NSF CAREER 1749742. We also gratefully acknowledge discussions with Dr. Evan Reed, Dr. Aaron Lindenberg, Dr. Jun-Ho Lee, and Dr. Young-Woo Son. D.Y.Q. and A.K. were supported by the U.S. Department of Energy (DOE), Office of Science, Office of Basic Energy Sciences Early Career Research Program under Award Number DE-SC0021965. The calculations used resources of the National Energy Research Scientific Computing (NERSC), a DOE Office of Science User Facility operated under contract no. DE-AC02-05CH11231; the Extreme Science and Engineering Discovery Environment (XSEDE), which is supported by National Science Foundation grant number ACI-1548562; and the Oak Ridge Leadership Computing Facility at the Oak Ridge National Laboratory, which is supported by the Office of Science of the U.S. DOE under Contract No. DE-AC05-00OR22725. We also thank the Yale Center for Research Computing for the calculations carried out on the Grace supercomputer.

**Methods**

*Device Fabrication*

The $WTe_2$ flakes were mechanically exfoliated from the bulk $WTe_2$ crystals purchased from 2D Semiconductors. $WTe_2$ flakes with desired lateral size and thickness were wet transferred onto $SiO_2$/Si wafer with the assistance of KOH[19]. The thickness of the flake was determined by Cypher ES atomic force microscope from Asylum Research. For the device fabrication, electrodes were written by electron beam lithography (Nabity NPGS, Helios G4 FIB-SEM) at a voltage of 30 kV and a current of 1.6 nA with a dose of 410 μC/cm$^2$. The developed devices were deposited with 15 nm Cr and 200 nm Au using a thermal evaporator (MBraun EcoVap) at a pressure of 10$^{-7}$ mbar. The devices were transferred into an argon glovebox with an O$_2$ and H$_2$O level below 1 ppm right after the liftoff to minimize the oxidation.

*Electrochemical Intercalation Cell Fabrication*

All the intercalation cells discussed in the main text adopted a planar cell configuration with both WTe$_2$ electrode and lithium electrode on a piece of transparent glass slide or SiO$_2$/Si wafer. The detailed fabrication steps are described in our previous research papers[19, 20]. For the liquid cells, the liquid electrolyte of 1 M LiPF$_6$ (EC/DEC, v:v= 50:50, battery-grade, Sigma-Aldrich) was injected into the cell to submerge the lithium and the device. For polymer cells, 0.227 g 1 M LiPF$_6$ solution was premixed with 0.475 g PEGMA and 1.145 g BEMA in a tarnished glass vial stirred overnight inside the glovebox. Then, 46 mg of photo-initiator was added to the well-mixed electrolyte and stirred for an additional hour in dark prior to the actual use. The mixed polymer electrolyte was drop-cast onto the target device using a micro-pipette covering both Li/copper foil and the WTe$_2$ device, then cured under UV light for 10 mins to form a gel-like electrolyte. The assembled intercalation cells have a cover slip with an air-tight seal that protects the cell from air and moisture.

*In situ Raman Characterization*

The assembled cell was taken out from the glovebox and quickly mounted to the Raman sample stage within 10 mins. Intercalation of lithium ions was driven by applying voltage between the Li and WTe$_2$ electrode potentiostatically with a Biological SP300 potentiostat/galvanostat. *In situ* Raman spectra were collected during lithium intercalation using a Horiba LabRAM HR Evolution Spectrometer (grating: 1800 lines/mm) with an excitation wavelength of 633 nm at 10% power (3.5 mW).

*In situ Low Frequency Raman and Angle-Resolved Raman Measurement*

The low frequency and angle dependent Raman spectra were acquired using a frequency-stabilized 785 nm laser (Toptica) as the excitation source with an incident power adjusted to 5.6 mW by a neutral-density filter. A set of five narrow-linewidth, reflective volume Bragg grating notch filters (OptiGrate) was used to reject the laser and allow for measurements of Raman signals down to about 5 cm$^{-1}$. The polarization of the incident beam was controlled by a zero-order half-waveplate, and the polarization of scattered Raman signal was first set by a broadband linear polarizer, and then rotated by a broadband half waveplate to maintain a fixed polarization of the Raman light sent into the spectrometer. The Raman signal was spatially filtered by a pair of 75-mm focal-length achromatic lens and a 50-µm pinhole, before sent into the spectrograph (Andor Kymera 328i) and captured by a Si EMCCD (Andor iXon Life 888). A super-long-working distance objective lens (10X, NA=0.28) was used in the measurements.

*In situ Two Terminal Electrical Transport Measurement*

*In situ* two terminal electrical transport data were collected using a semiconductor device analyzer (Agilent Technologies B1500A) during the potentiostatic lithium intercalation. The drain-source voltage applied was 50 mV, and the resistance was extracted from the slope of the *I-V* curve using a linear regression method.

*Low Temperature Magneto-Transport Measurement*

Magneto-transport measurements of pristine $T_d$-$WTe_2$ at 2 K were carried out with Quantum Design's DynaCool PPMS system at AC mode (17.777 Hz). The system was first cooled to 2 K at 0 T with a cooling rate of 5 K/min. Then the magneto-transport measurement of the pristine state was carried out with magnetic field sweeping between -14 to 14 T at a sweeping rate of 100 Oe/s. Two channel fitting model ($\rho_{xy} = \frac{(n_h\mu_h^2 - n_e\mu_e^2)B + \mu_h^2\mu_e^2(n_h - n_e)B^3}{e[(n_h\mu_h + n_e\mu_e)^2 + (n_h - n_e)^2\mu_h^2\mu_e^2 B^2]}$; $\rho_{xx} = \frac{(n_h\mu_h + n_e\mu_e) + (n_h\mu_h\mu_e^2 + n_e\mu_e\mu_h^2)B^2}{e[(n_h\mu_h + n_e\mu_e)^2 + (n_h - n_e)^2\mu_h^2\mu_e^2 B^2]}$ ) was applied to fit the magnetic field dependent $R_{xx}$ and $R_{xy}$ to extract out the mobility ($\mu$) and carrier density ($n$) of electron ($e$) and hole ($h$)[17].

*In situ Magneto-Transport Measurement*

*In situ* Hall transport measurements were carried out using the DynaCool PPMS system. After the assembly of the polymer cell and wire bonding between the gold pads of the PPMS puck and the copper foil of the polymer cell, the puck was quickly transferred into the chamber of the PPMS system. Intercalation was carried out at 300 K and 3 T, and $R_{xx}$ and $R_{xy}$ were recorded during the intercalation. The carrier density was determined by equations $n_H = B/(eR_{xy})$, $e = 1.602 \times 10^{-19} C$ [21]. The correction of $R_{xy}$ is obtained by subtracting the contribution of $R_{xx}$ from raw $R_{xy}$ using the equation $R_{xy,corrected,3T} = R_{xy,raw,3T} - \alpha \cdot R_{xx,3T}$, where $\alpha$ is a geometric factor determined by the ratio of raw $R_{xy}$ over $R_{xx}$ at zero field using $\alpha = R_{xy,0T}/R_{xx,0T}$. Temperature-dependent resistance measurements were carried out by first freezing the electrochemical intercalation through a rapid cooling to 200 K at a rate of 10 K/min, at which point the intercalation current running through the cell gradually goes to zero. Then the system was gradually cooled to 150 K at a rate of 2 K/min.

*Post-mortem characterizations*

Intercalation devices were disassembled by physically removing the liquid or polymer electrolyte by a razor blade. Then the device was immersed in isopropanol to further remove the residual electrolyte. The cleaned device was characterized by Raman spectroscopy (Horiba LabRAM HR Evolution Spectrometer; 633 nm) and SEM (Helios G4 FIB-SEM) with a stage tilting angle of 50° at a voltage of 5 kV and a current of 25 pA.

*Ab initio Calculations*

Density functional theory (DFT) calculations were carried out in a plane-wave basis set within the Projector Augmented Wave (PAW) approach [22, 23] as implemented in the Quantum Espresso software package[24]. The exchange-correlation was treated at the Generalized Gradient Approximation (GGA) level of Perdew, Burke, and Ernzehof (PBE) [25]. The van der Waals interactions were accounted for using Grimme's D3 correction with Becke-Johnson damping [26]. A kinetic energy cut-off of 1040 eV was used for the expansion of the plane-wave basis in all the calculations, and the Brillouin zone was sampled using Gamma-centered Monkhorst-Pack [27] k-point meshes of 32×16×8 and 16×8×8 for the $T_d$ (12 atoms) and $T_d'$ (24 atoms) unit cells respectively. The unit cells were relaxed until the total energy and the force on each atom converged to within 0.02 meV/atom and 0.03 eV/Å respectively. The lithiated $T_d$ phase was relaxed with the volume fixed to that of the pristine $T_d$ to prevent a doping-induced phase transition. Spin-orbit coupling with non-colinear magnetization was used for all the band structure calculations. Phonopy [28] was used for the calculation of phonon bands, with supercell sizes of 2×2×2 and 2×2×1 for $T_d$- and $T_d'$- WTe$_2$ phases, and q-meshes of 8×4×2 and 4×4×4 were used respectively. For implicitly doped $T_d$, a 2×2×1 supercell with a q-mesh of 4×4×4 was used.

**References**

1. Fatemi, V.; Wu, S.; Cao, Y.; Bretheau, L.; Gibson, Q. D.; Watanabe, K.; Taniguchi, T.; Cava, R. J.; Jarillo-Herrero, P., Electrically tunable low-density superconductivity in a monolayer topological insulator. *Science* **2018,** *362* (6417), 926.
2. Sajadi, E.; Palomaki, T.; Fei, Z.; Zhao, W.; Bement, P.; Olsen, C.; Luescher, S.; Xu, X.; Folk, J. A.; Cobden, D. H., Gate-induced superconductivity in a monolayer topological insulator. *Science* **2018,** *362* (6417), 922.
3. Tang, S.; Zhang, C.; Wong, D.; Pedramrazi, Z.; Tsai, H.-Z.; Jia, C.; Moritz, B.; Claassen, M.; Ryu, H.; Kahn, S.; Jiang, J.; Yan, H.; Hashimoto, M.; Lu, D.; Moore, R. G.; Hwang, C.-C.; Hwang, C.; Hussain, Z.; Chen, Y.; Ugeda, M. M.; Liu, Z.; Xie, X.; Devereaux, T. P.; Crommie, M. F.; Mo, S.-K.; Shen, Z.-X., Quantum spin Hall state in monolayer 1T'-WTe$_2$. *Nature Physics* **2017,** *13* (7), 683-687.
4. Ono, M.; Noji, T.; Harada, M.; Sato, K.; Kawamata, T.; Kato, M., New Lithium- and Ethylenediamine-Intercalated Superconductors Li$_x$(C$_2$H$_8$N$_2$)$_y$WTe$_2$. *Journal of the Physical Society of Japan* **2020,** *90* (1), 014706.
5. Zhu, L.; Li, Q.-Y.; Lv, Y.-Y.; Li, S.; Zhu, X.-Y.; Jia, Z.-Y.; Chen, Y. B.; Wen, J.; Li, S.-C., Superconductivity in Potassium-Intercalated T$_d$-WTe$_2$. *Nano Letters* **2018,** *18* (10), 6585-6590.
6. Nayak, A. K.; Steinbok, A.; Roet, Y.; Koo, J.; Margalit, G.; Feldman, I.; Almoalem, A.; Kanigel, A.; Fiete, G. A.; Yan, B.; Oreg, Y.; Avraham, N.; Beidenkopf, H., Evidence of topological boundary modes with topological nodal-point superconductivity. *Nature Physics* **2021**.
7. Sipos, B.; Kusmartseva, A. F.; Akrap, A.; Berger, H.; Forró, L.; Tutiš, E., From Mott state to superconductivity in 1T-TaS$_2$. *Nature Materials* **2008,** *7* (12), 960-965.
8. Yu, Y.; Yang, F.; Lu, X. F.; Yan, Y. J.; Cho, Y.-H.; Ma, L.; Niu, X.; Kim, S.; Son, Y.-W.; Feng, D.; Li, S.; Cheong, S.-W.; Chen, X. H.; Zhang, Y., Gate-tunable phase transitions in thin flakes of 1T-TaS$_2$. *Nature Nanotechnology* **2015,** *10* (3), 270-276.
9. Ye, J. T.; Zhang, Y. J.; Akashi, R.; Bahramy, M. S.; Arita, R.; Iwasa, Y., Superconducting Dome in a Gate-Tuned Band Insulator. *Science* **2012,** *338* (6111), 1193.
10. Lee, J.-H.; Son, Y.-W., Gate-tunable superconductivity and charge-density wave in monolayer 1T′-MoTe$_2$ and 1T′-WTe$_2$. *Physical Chemistry Chemical Physics* **2021,** *23* (32), 17279-17286.
11. Yang, W.; Mo, C.-J.; Fu, S.-B.; Yang, Y.; Zheng, F.-W.; Wang, X.-H.; Liu, Y.-A.; Hao, N.; Zhang, P., Soft-Mode-Phonon-Mediated Unconventional Superconductivity in Monolayer 1T'-WTe$_2$. *Physical Review Letters* **2020,** *125* (23), 237006.
12. Marini, G.; Calandra, M., Light-tunable charge density wave orders in MoTe$_2$ and WTe$_2$ single layers. *arXiv preprint arXiv:2111.11920* **2021**.

13. Muscher, P. K.; Rehn, D. A.; Sood, A.; Lim, K.; Luo, D.; Shen, X.; Zajac, M.; Lu, F.; Mehta, A.; Li, Y.; Wang, X.; Reed, E. J.; Chueh, W. C.; Lindenberg, A. M., Highly Efficient Uniaxial In-Plane Stretching of a 2D Material via Ion Insertion. *Advanced Materials* **2021,** *33*, 2101875.
14. Kim, M.; Han, S.; Kim, J. H.; Lee, J.-U.; Lee, Z.; Cheong, H., Determination of the thickness and orientation of few-layer tungsten ditelluride using polarized Raman spectroscopy. *2D Materials* **2016,** *3* (3), 034004.
15. Eda, G.; Yamaguchi, H.; Voiry, D.; Fujita, T.; Chen, M.; Chhowalla, M., Photoluminescence from Chemically Exfoliated $MoS_2$. *Nano Letters* **2011,** *11* (12), 5111-5116.
16. Ali, M. N.; Xiong, J.; Flynn, S.; Tao, J.; Gibson, Q. D.; Schoop, L. M.; Liang, T.; Haldolaarachchige, N.; Hirschberger, M.; Ong, N. P.; Cava, R. J., Large, non-saturating magnetoresistance in $WTe_2$. *Nature* **2014,** *514* (7521), 205-208.
17. Woods, J. M.; Shen, J.; Kumaravadivel, P.; Pang, Y.; Xie, Y.; Pan, G. A.; Li, M.; Altman, E. I.; Lu, L.; Cha, J. J., Suppression of Magnetoresistance in Thin $WTe_2$ Flakes by Surface Oxidation. *ACS Applied Materials & Interfaces* **2017,** *9* (27), 23175-23180.
18. Tang, W.; Sanville, E.; Henkelman, G., A grid-based Bader analysis algorithm without lattice bias. *Journal of Physics: Condensed Matter* **2009,** *21* (8), 084204.
19. Yazdani, S.; Pondick, J. V.; Kumar, A.; Yarali, M.; Woods, J. M.; Hynek, D. J.; Qiu, D. Y.; Cha, J. J., Heterointerface Effects on Lithium-Induced Phase Transitions in Intercalated $MoS_2$. *ACS Applied Materials & Interfaces* **2021,** *13* (8), 10603-10611.
20. Pondick, J. V.; Kumar, A.; Wang, M.; Yazdani, S.; Woods, J. M.; Qiu, D. Y.; Cha, J. J., Heterointerface Control over Lithium-Induced Phase Transitions in MoS2 Nanosheets: Implications for Nanoscaled Energy Materials. *ACS Applied Nano Materials* **2021**.
21. Bediako, D. K.; Rezaee, M.; Yoo, H.; Larson, D. T.; Zhao, S. Y. F.; Taniguchi, T.; Watanabe, K.; Brower-Thomas, T. L.; Kaxiras, E.; Kim, P., Heterointerface effects in the electrointercalation of van der Waals heterostructures. *Nature* **2018,** *558* (7710), 425-429.
22. Blöchl, P. E., Projector augmented-wave method. *Physical Review B* **1994,** *50* (24), 17953-17979.
23. Dal Corso, A., Pseudopotentials periodic table: From H to Pu. *Computational Materials Science* **2014,** *95*, 337-350.
24. Giannozzi, P.; Baroni, S.; Bonini, N.; Calandra, M.; Car, R.; Cavazzoni, C.; Ceresoli, D.; Chiarotti, G. L.; Cococcioni, M.; Dabo, I., QUANTUM ESPRESSO: a modular and open-source software project for quantum simulations of materials. *Journal of physics: Condensed matter* **2009,** *21* (39), 395502.
25. Perdew, J. P.; Burke, K.; Ernzerhof, M., Generalized gradient approximation made simple. *Physical review letters* **1996,** *77* (18), 3865.
26. Becke, A. D.; Johnson, E. R., A density-functional model of the dispersion interaction. *The Journal of chemical physics* **2005,** *123* (15), 154101.
27. Monkhorst, H. J.; Pack, J. D., Special points for Brillouin-zone integrations. *Physical Review B* **1976,** *13* (12), 5188-5192.
28. Togo, A.; Tanaka, I., First principles phonon calculations in materials science. *Scripta Materialia* **2015,** *108*, 1-5.
29. Mehl, M. J.; Hicks, D.; Toher, C.; Levy, O.; Hanson, R. M.; Hart, G.; Curtarolo, S., The AFLOW library of crystallographic prototypes: part 1. *Computational Materials Science* **2017,** *136*, S1-S828.

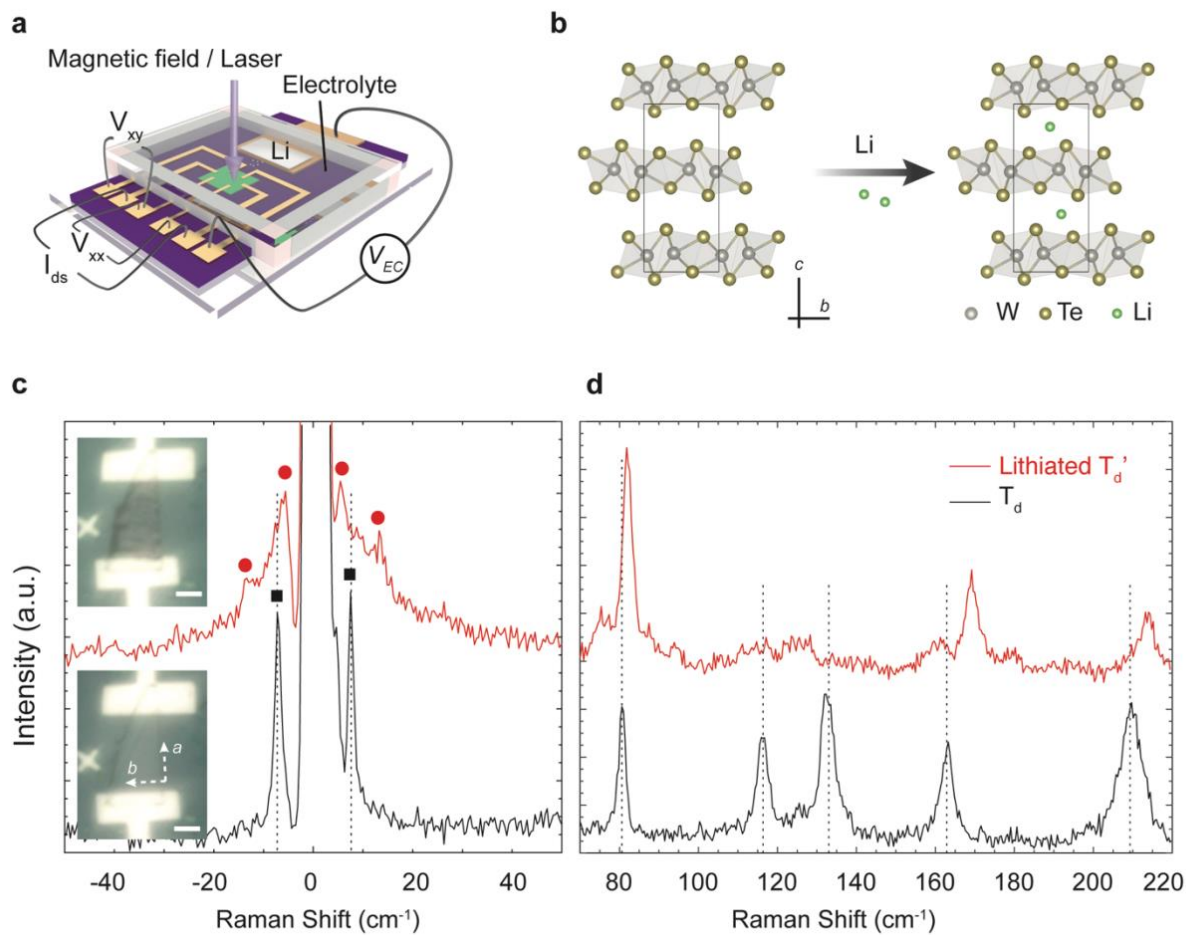

**Figure 1.** Phase transition in $T_d$-$WTe_2$ induced by lithium intercalation. **(a).** Schematic of a lithium-ion electrochemical cell with capabilities of *in situ* Raman and *in situ* Hall measurement. $V_{EC}$ represents the electrochemical intercalation voltage applied between the Li (silver rectangle) and $WTe_2$ (green rectangle). **(b).** Side-view of lithium intercalation in the van der Waals gaps of $T_d$-$WTe_2$ (W: grey, Te: gold, Li: green). **(c-d).** Low frequency (**c**) and high frequency (**d**) Raman spectra of the $T_d$ phase (black, bottom) and lithiated $T_d'$ phase (red, top). Insets in **c** are the corresponding optical images of the $WTe_2$ device before (bottom) and after (top) the phase transition. Scale bars are 10 μm. Red circles and black squares in **c** mark the positions of Raman active modes in $T_d$ and the new phase, respectively. Black vertical dashed lines in **c** and **d** are visual guides for the positions of Raman modes in $T_d$ phase.

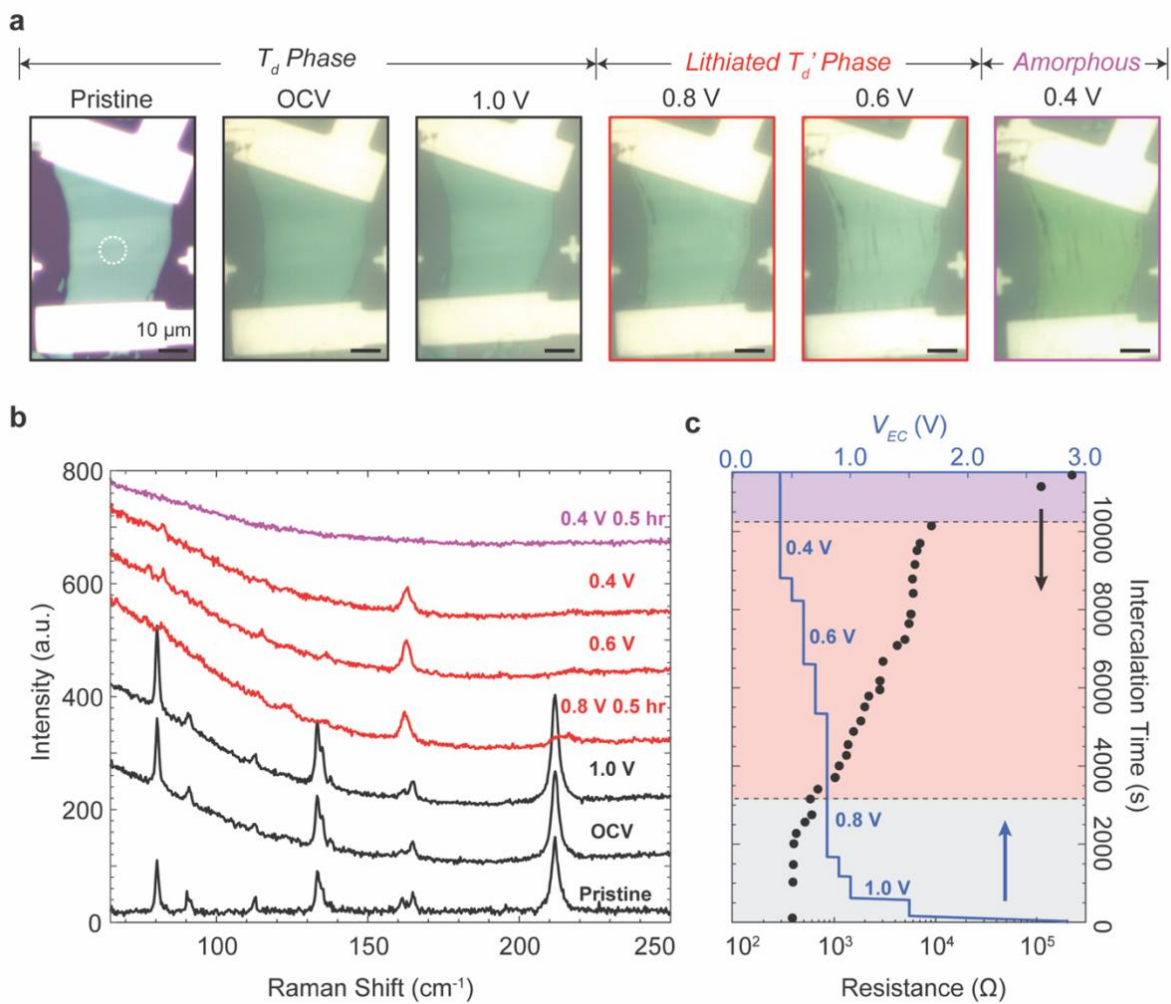

**Figure 2.** *In situ* optical characterization and two terminal electrical transport measurements of a WTe$_2$ device as a function of electrochemical intercalation voltage. **(a)**. Optical images of a WTe$_2$ device as a function of intercalation driving voltage. According to the optical contrast, the intercalation consists of three stages: T$_d$ phase (black framed), new phase (red framed), and amorphous state (purple framed). The white dashed circle denotes the position for *in situ* Raman data collection. **(b).** Stacking plot of high frequency Raman spectra as a function of intercalation voltage with the color scheme corresponding to the phase separation in **(a)**. The background of the Raman spectra from OCV to 0.4 V comes from the polymer electrolyte. **(c)**. Simultaneous two terminal resistance (solid black circles) as function of intercalation time with the intercalation direction lined up with the Raman stacking sequence in **(b)**. The blue line is the profile of electrochemical intercalation voltage as a function of time. In line with **(a)** and **(b)**, three phases are divided and colored correspondingly.

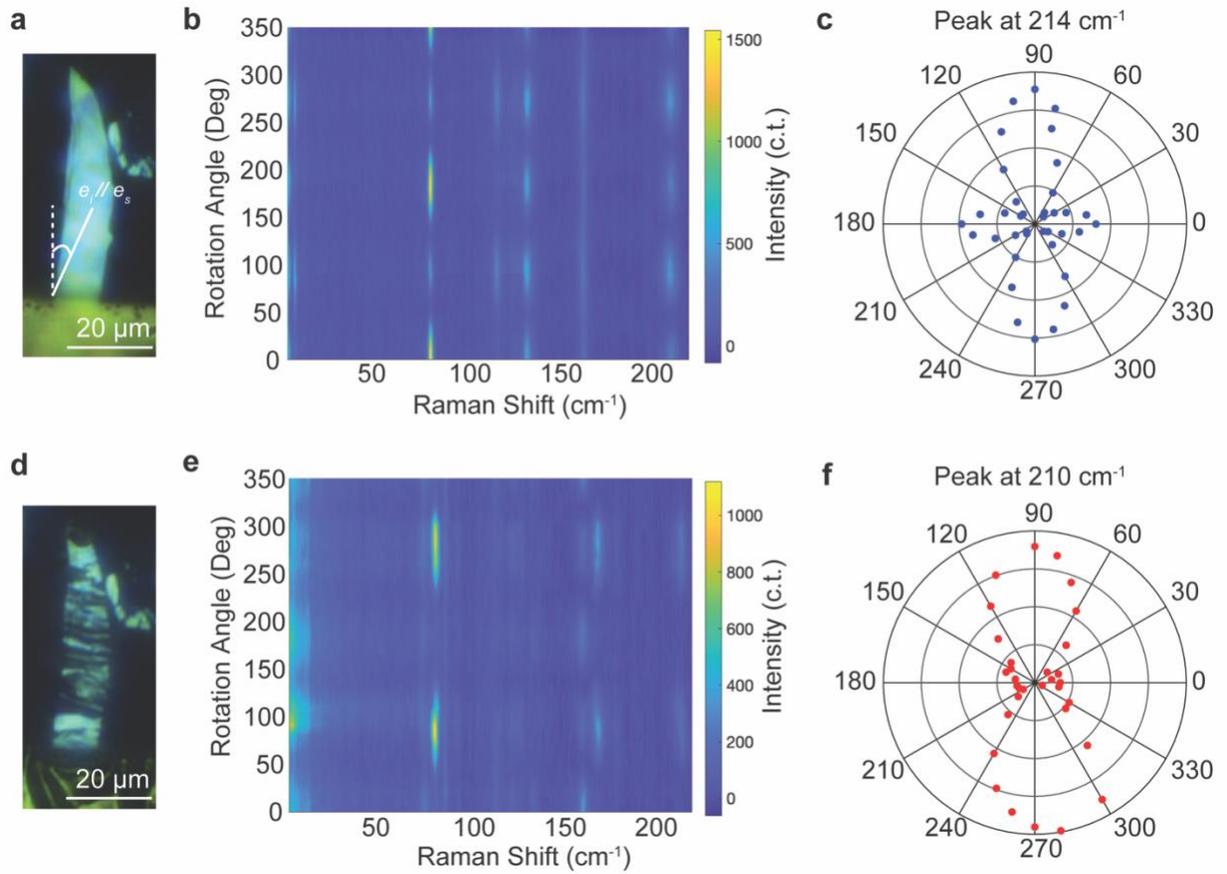

**Figure 3**. Angle-resolved Raman characterization of $T_d$ (**a-c**) and lithiated $T_d'$-WTe$_2$ (**d-f**). (**a**) and (**d**) are optical images of the WTe$_2$ flake at pristine $T_d$ (**a**) and lithiated $T_d'$ (**d**) phase, respectively. The angle is defined as the angle between the vertical white dashed line and the polarization direction of the incident/scattered beam in (**a**). (**b**) and (**e**) are the angle-resolved Raman spectral maps of the $T_d$ (**b**) and lithiated $T_d'$ (**e**) phase, respectively. (**c**) and (**f**) are the pole graphs of the Raman peak centered around 210 cm$^{-1}$ in $T_d$ (**c**) and lithiated $T_d'$ (**f**) phase, respectively.

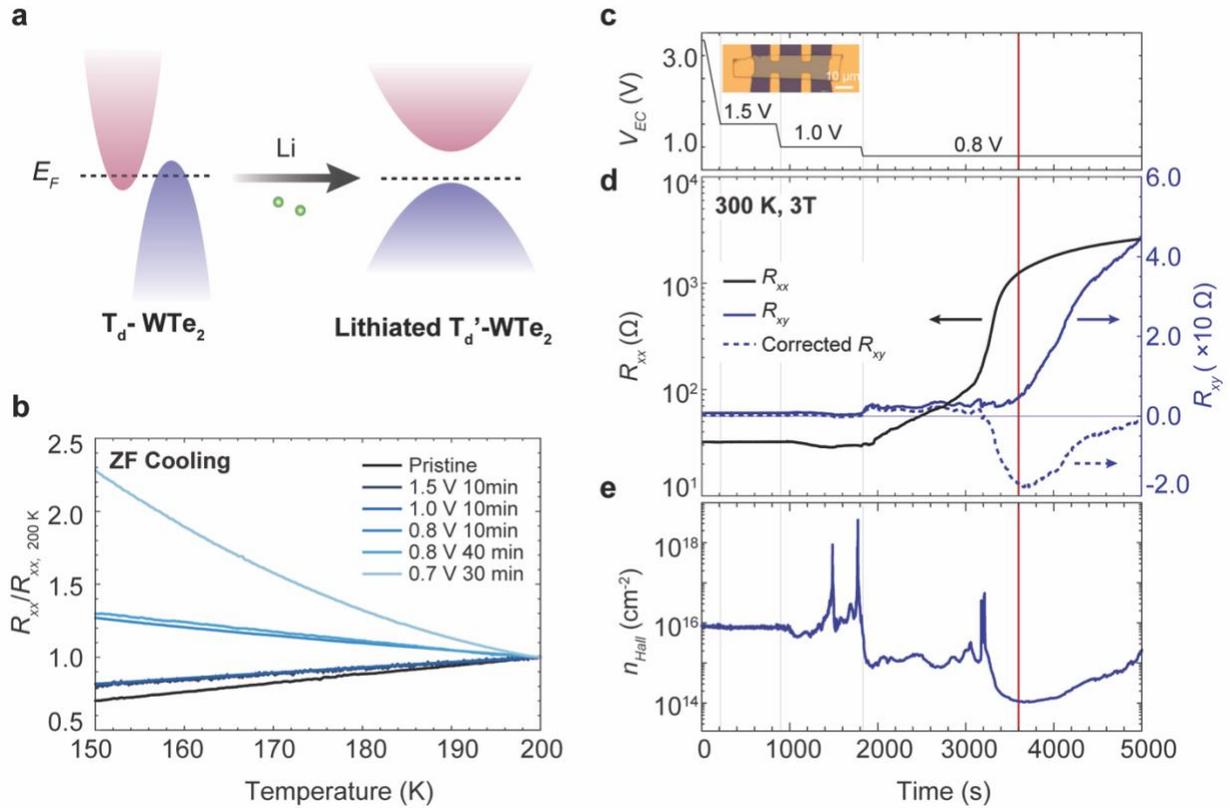

**Figure 4.** *In situ* Hall measurement of a WTe$_2$ device as a function of lithiation at 300 K. **(a).** Schematic of electronic band structure changes during the phase transition from T$_d$-WTe$_2$ to lithiated T$_d$'-WTe$_2$. **(b).** Zero field (ZF) cooling curves of $R_{xx}$ at pristine state (black) and various stages of intercalation (blue). A limited temperature window from 200 K to 150 K was due to the polymer electrolyte instability below 150 K. **(c).** Profile of electrochemical intercalation voltage as a function of intercalation time. **(d).** Raw data of $R_{xx}$ (black) and $R_{xy}$ (blue), and corrected $R_{xy}$ (blue dashed, see Methods) as a function of intercalation time at 3 T and 300 K. The grey horizontal line marks the position of zero for $R_{xy}$. Inset is the optical image of a WTe$_2$ Hall bar device. **(e).** Carrier density determined from corrected $R_{xy}$ as a function of intercalation time. The light grey vertical lines in (**c-e**) mark the time of applying each new intercalation voltage. The dark red vertical line in (**c-e**) denotes the time when new Raman modes emerge.

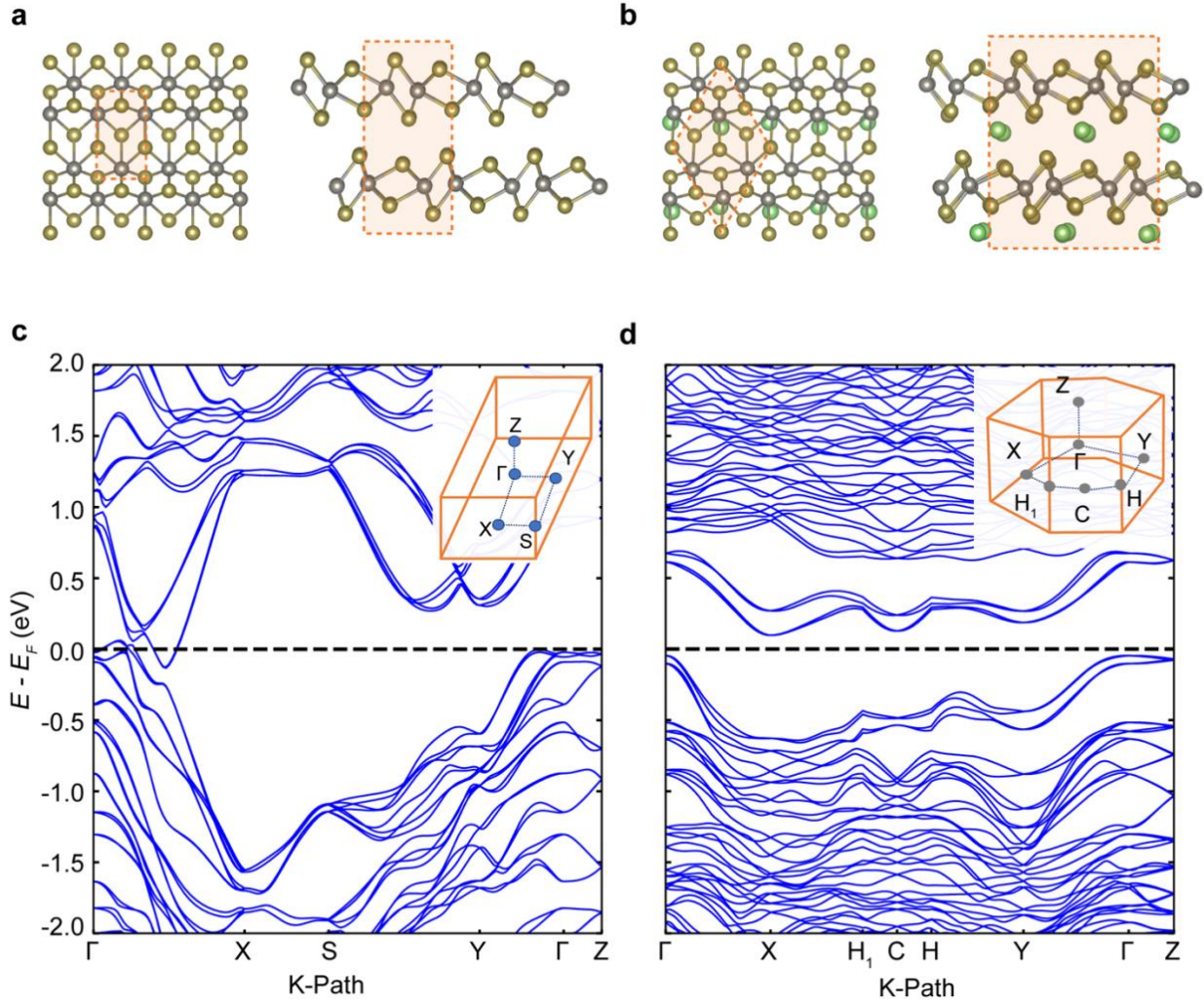

**Figure 5.** In-plane and cross-sectional atomic structures of $T_d$-WTe$_2$ **(a)** and lithiated $T_d'$-Li$_{0.5}$WTe$_2$ **(b)** with W atoms in grey, Te atoms in gold, and Li atoms in green. Unit cells are highlighted with orange dashed lines. DFT electronic band structures of **(c)** $T_d$-WTe$_2$ and **(d)** lithiated $T_d'$- Li$_{0.5}$WTe$_2$ along a high-symmetry K-path in Brillouin zone, as shown in the inset[29], reveal their semi-metallic and semiconducting nature, respectively. The horizontal black dashed line in **(c)** and **(d)** marks the Fermi level for each phase.

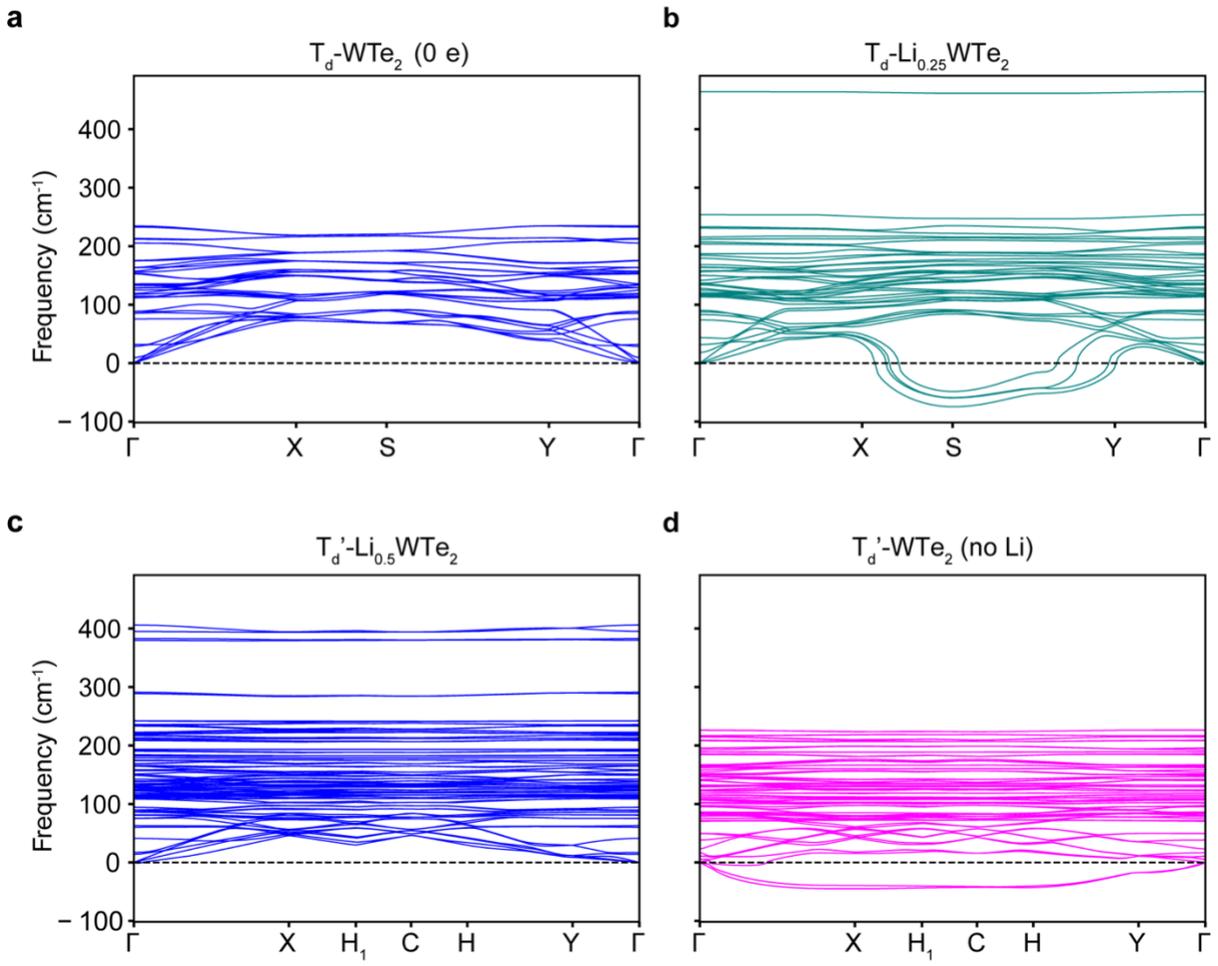

**Figure 6.** Phonon-dispersion of (**a**) T$_d$-WTe$_2$, (**b**) lithiated T$_d$-Li$_{0.25}$WTe$_2$, (**c**) lithiated T$_d$'-Li$_{0.5}$WTe$_2$ phase, and (**d**) T$_d$'-WTe$_2$ without any Li atoms.